\documentclass[11pt,fleqn]{article}
\usepackage[utf8]{inputenc}
\usepackage[top=1in,bottom=1in,left=1in,right=1in]{geometry}
\usepackage{fouriernc}
\usepackage{authblk}
\usepackage{mathtools}
\usepackage{multicol}
\usepackage{amssymb}
\newcommand{\ket}[1]{\rvert#1\rangle}
\newcommand{\bra}[1]{\langle #1\rvert}

\newcommand{\bbraket}[4]{{}_{#1}\langle #2\rvert#3\rangle_{#4}}
\newcommand{\kket}[2]{\rvert#1\rangle_{#2}}

\newcommand{\Braket}[3]{\bra{#1}#2\ket{#3}}
\newcommand{\BBraket}[5]{{}_{#1}\bra{#2}\;#3\;\ket{#4}_{#5}}
\newcommand{\qbinom}[2]{\genfrac{[}{]}{0pt}{}{#1}{#2}_q}
\newcommand*\pfqskip{8mu}
\catcode`,\active
\newcommand*\pfq{\begingroup
        \catcode`\,\active
        \def ,{\mskip\pfqskip\relax}%
        \dopfq
}
\catcode`\,12
\def\dopfq#1#2#3#4#5{%
        {}_{#1}\phi_{#2}\left(\genfrac..{0pt}{}{#3}{#4}\,\Big\rvert\,#5\right)%
        \endgroup
}
\newcommand*\pFqskip{8mu}
\catcode`,\active
\newcommand*\pFq{\begingroup
        \catcode`\,\active
        \def ,{\mskip\pFqskip\relax}%
        \dopFq
}
\catcode`\,12
\def\dopFq#1#2#3#4#5{%
        {}_{r}\phi_{s}\bigg(\genfrac..{0pt}{}{#3}{#4}\,\Big\rvert\,#5\biggr)%
        \endgroup
}
\newcommand*\pGqskip{8mu}
\catcode`,\active
\newcommand*\pGq{\begingroup
        \catcode`\,\active
        \def ,{\mskip\pGqskip\relax}%
        \dopGq
}
\catcode`\,12
\def\dopGq#1#2#3#4#5{%
        {}_{#1}F_{#2}\bigg(\genfrac..{0pt}{}{#3}{#4}\,\Big\rvert\,#5\biggr)%
        \endgroup
}
\title{$q$-Rotations and Krawtchouk polynomials}
\author[1]{Vincent X. Genest}
\author[2]{Sarah Post}
\author[1]{Luc Vinet}
\author[3]{Guo-Fu Yu}
\author[4]{Alexei Zhedanov}
\affil[1]{Centre de recherches math\'ematiques, Universit\'e de Montr\'eal, P.O. Box 6128, Centre-ville Station, Montr\'eal, Canada, H3C 3J7}
\affil[2]{Department of Mathematics, University of Hawai'i, Honolulu, HI 96822, USA}
\affil[3]{Department of Mathematics, Shanghai Jiao Tong University, Shanghai 200240, China}
\affil[4]{Donetsk Institute for Physics and Technology, Donetsk 340114, Ukraine}
\date{}

\begin{document}
\maketitle
\thispagestyle{empty}
\hrule
\begin{abstract}
\noindent
An algebraic interpretation of the one-variable quantum $q$-Krawtchouk polynomials is provided in the framework of the Schwinger realization of $\mathcal{U}_{q}(sl_{2})$ involving two independent $q$-oscillators. The polynomials are shown to arise as matrix elements of unitary ``$q$-rotation'' operators expressed as $q$-exponentials in the $\mathcal{U}_{q}(sl_{2})$ generators. The properties of the polynomials (orthogonality relation, generating function, structure relations, recurrence relation, difference equation) are derived by exploiting the algebraic setting. The results are extended to another family of polynomials, the affine $q$-Krawtchouk polynomials, through a duality relation.
\bigskip

\\
\noindent \textbf{AMS classification numbers:} 33D45, 16T05 
\end{abstract}
\hrule
\section{Introduction}
This paper is concerned with the algebraic interpretation and characterization of two families of univariate basic orthogonal polynomials: the quantum and the affine $q$-Krawtchouk polynomials. 

Recently, some of us have offered \cite{Genest-2013-06, Genest-2013-07-2} a remarkably simple description of the multivariate Krawtchouk orthogonal polynomials introduced by Griffiths \cite{Griffiths-1971-04} as matrix elements of the unitary representations of the orthogonal groups on multi-oscillator quantum states. These polynomials of Griffiths have as special cases the polynomials of Krawtchouk type proposed by Tratnik \cite{Tratnik-1991-04} which correspond to particular rotations. $q$-analogs of these Tratnik polynomials were offered by Gasper and Rahman \cite{Gasper-2007}; their bispectrality was established by Geronimo and Iliev \cite{Geronimo-2010} in the $q=1$ case and by Iliev \cite{Iliev-2011} in the basic case. It would be desirable to obtain the algebraic underpinning of the multivariate $q$-Krawtchouk polynomials that parallels the fruitful framework developed in the $q=1$ case, that is to relate the polynomials to matrix elements of ``$q$-rotations'' on $q$-oscillator states.

As a first essential step towards that goal, we elaborate here this picture in the one-variable case. The univariate quantum $q$-Krawtchouk polynomials will be shown to arise as matrix elements of products of $q$-exponentials in $\mathcal{U}_{q}(sl_2)$ generators realized \`a la Schwinger with two independent $q$-oscillators. By conjugation, these operators effect non-linear automorphisms of the quantum algebra $\mathcal{U}_{q}(sl_{2})$. While this connection between $\mathcal{U}_q(sl_{2})$ and $q$-analogs of the Krawtchouk polynomials is mentioned in \cite{Zhedanov-1993-06}, the detailed characterization needed for an extension to an arbitrary number of variables is carried out here in full. Using the algebraic interpretation, the main properties of the quantum $q$-Krawtchouk polynomials such as the orthogonality relation, the generating function, the structure relations, the difference equation, and the recurrence relation will be derived. Our approach will be seen to entail similar results for the univariate affine $q$-Krawtchouk polynomials through a duality relation. The novelty of the results presented here does not lie of course in the characteristic formulas for the polynomials but in their detailed algebraic interpretation.

Let us point out that an algebraic interpretation of the $q$-Krawtchouk was originally obtained by Koornwinder some time ago \cite{Koornwinder-1989} in a quantum group setting, that is using the quantum function algebra dual to $\mathcal{U}_{q}(sl_2)$. A comparison and a detailed connection between the quantum group approach and the quantum algebra one favored here was given in \cite{Floreanini-1993}. While the two methods are fundamentally equivalent, the latter closely follows the representations of Lie groups via the exponentiation of algebra generators and reveals simplicity advantages that shall be helpful in higher dimensional generalizations. In that vein, the embedding of $\mathcal{U}_{q}(sl_2)$ in the two-dimensional $q$-Weyl algebra via the use of $q$-oscillators and the Schwinger realization offers a refined structure that entails, as shall be seen, forward and backward relations for the polynomials. Let us further note that a number of other different algebraic treatments of the various $q$-generalizations of the Krawtchouk polynomials can be found in \cite{Atakishiyev-2004, Delsarte-1976,Koelink-2000,Smirnov-2004,Stanton-1981}.

The paper is organized as follows. In Section I, some elements of $q$-analysis are reviewed, the Schwinger realization of $\mathcal{U}_q(sl_2)$ is revisited and the unitary $q$-rotation operators are constructed. In section II, the matrix elements of the $q$-rotation operators are calculated directly and expressed in terms of the quantum $q$-Krawtchouk polynomials. In section III, the structure relations for the polynomials are derived. In section IV, two types of generating functions are obtained. In section VI, the recurrence relation and the difference equation are recovered. In section VII, the duality relation between the quantum $q$-Krawtchouk and the affine $q$-Krawtchouk is examined. A conclusion follows.

\section{The Schwinger model for $\mathcal{U}_q(sl_2)$ and $q$-rotations}
In this section, the necessary elements of $q$-analysis are presented, the Schwinger realization of $\mathcal{U}_q(sl_2)$ is reviewed, and the unitary $q$-rotation operators are constructed.
\subsection{Elements of $q$-analysis}
We adopt the notation and conventions of \cite{Gasper-2004}. The basic hypergeometric series is defined by 
\begin{align}
\label{Basic-Hyper}
\pFq{3}{3}{a_1,\ldots,a_{r}}{b_1,\ldots,b_{s}}{q,z}=\sum_{n=0}^{\infty}\frac{(a_1;q)_{n}\cdots (a_r;q)_{n}}{(q;q)_{n}(b_1;q)_{n}\cdots (b_s;q)_{n}}\,\left[(-1)^{n}q^{\binom{n}{2}}\right]^{1+s-r}\,z^{n},
\end{align}
with $\binom{n}{2}=n(n-1)/2$ and where $(a;q)_n$ stands for the $q$-shifted factorial
\begin{align*}
(a;q)_{n}=
\begin{cases}
1, & n=0,
\\
(1-a)(1-aq)\cdots (1-aq^{n-1}), & n=1,2,\ldots.
\end{cases}
\end{align*}
The $q$-shifted factorials satisfy a number of identities (see Appendix I of \cite{Gasper-2004}); for example, a direct expansion shows that
\begin{align}
\label{Identity-1}
(a;q)_{n-k}=\frac{(a;q)_{n}}{(q^{1-n}/a;q)_k}\,(-q/a)^{k}\,q^{\binom{k}{2}-nk},
\end{align}
where $n$ and $k$ are integers. The $q$-binomial coefficients are defined by
\begin{align}
\label{QBinom}
\qbinom{a}{b}=\frac{(q;q)_{a}}{(q;q)_{b}(q;q)_{a-b}}.
\end{align}
It is seen that in the limit $q\uparrow 1$, the coefficients \eqref{QBinom} tend to the ordinary binomial coefficients. The $q$-exponential functions will play an important role in what follows. The little $q$-exponential, denoted by $e_{q}(z)$, is defined as
\begin{align}
\label{Little-Exp}
e_q(z)=\pfq{1}{0}{0}{-}{q,z}=\sum_{n=0}^{\infty}\frac{z^{n}}{(q;q)_{n}}=\frac{1}{(z;q)_{\infty}},
\end{align}
for $|z|<1$ and the big $q$-exponential, denoted $E_{q}(z)$, is given by
\begin{align}
\label{Big-Exp}
E_{q}(z)=\pfq{0}{0}{-}{-}{q,-z}=\sum_{n=0}^{\infty}\frac{q^{\binom{n}{2}}}{(q;q)_{n}}z^{n}=(-z;q)_{\infty}.
\end{align}
It follows that $e_{q}(z)E_{q}(-z)=1$. The $q$-extensions of the Baker-Campbell-Hausdorff formula \cite{Vinet-1993,Kalnins-1993} shall be needed. The first relation reads
\begin{align}
\label{Conjugation-1}
E_{q}(\lambda X)Ye_{q}(-\lambda X)=\sum_{n=0}^{\infty}\frac{\lambda^{n}}{(q;q)_{n}}[X,Y]_{n},
\end{align}
where 
\begin{align*}
[X,Y]_{0}=Y,\qquad [X,Y]_{n+1}=q^{n}\,X[X,Y]_{n}-[X,Y]_{n}X,\qquad n=0,1,2,\ldots
\end{align*}
The second relation is of the form
\begin{align}
\label{Conjugation-2}
e_{q}(\lambda X)YE_{q}(-\lambda X)=\sum_{n=0}^{\infty}\frac{\lambda^{n}}{(q;q)_{n}}[X,Y]_{n}',
\end{align}
where 
\begin{align*}
[X,Y]_{0}'=Y,\qquad [X,Y]_{n+1}'=X[X,Y]_{n}'-q^{n}\,[X,Y]_{n}'X,\qquad n=0,1,2,\ldots
\end{align*}
One has also the identities
\begin{align}
\label{Q-Commutation}
e_{q}(X+Y)=e_{q}(Y)e_{q}(X),\qquad \text{and} \qquad E_{q}(X+Y)=E_{q}(X)E_{q}(Y),
\end{align}
for $XY=q YX$.
\subsection{The Schwinger model for $\mathcal{U}_q(sl_{2})$}
Consider two mutually commuting sets  $\{A_{\pm}, A_0\}$ and $\{B_{\pm},B_0\}$ of  $q$-oscillator algebra generators that satisfy the commutation relations
\begin{subequations}
\label{Algebra}
\begin{alignat}{3}
[A_0,A_{\pm}]&=\pm A_{\pm},&\qquad [A_{-},A_{+}]&=q^{A_0},\qquad A_{-}A_{+}-q A_{+}A_{-}&=1,
\\
[B_0,B_{\pm}]&=\pm B_{\pm},&\qquad [B_{-},B_{+}]&=q^{B_0},\qquad B_{-}B_{+}-q B_{+}B_{-}&=1,
\end{alignat}
\end{subequations}
and $[A_{\cdot},B_{\cdot}]=0$. It follows from \eqref{Algebra} that 
\begin{align*}
A_{+}A_{-}=\frac{1-q^{A_0}}{1-q},\qquad B_{+}B_{-}=\frac{1-q^{B_0}}{1-q}.
\end{align*}
The algebra \eqref{Algebra} has a standard representation on the orthonormal states
\begin{align}
\label{States}
\ket{n_{A},n_{B}}\equiv \ket{n_{A}}\otimes \ket{n_{B}},\qquad n_A,n_{B}=0,1,2,\ldots
\end{align}
defined by the following actions of the generators on the factors of the tensor product states:
\begin{align}
\label{Rep}
X_{-}\ket{n_X}=\sqrt{\frac{1-q^{n_x}}{1-q}}\,\ket{n_X-1},\quad X_{+}\ket{n_X}=\sqrt{\frac{1-q^{n_X+1}}{1-q}}\ket{n_X+1},\quad X_0\ket{n_X}=n_X\ket{n_X},
\end{align}
with $X=A$ or $B$. It is seen that when $q\uparrow 1$, the representation \eqref{Rep} goes to the standard oscillator representation (see for example Chap. 5 of \cite{Cohen-1973}). Moreover, one has $X_{\pm}^{\dagger}=X_{\mp}$ in this representation. The Schwinger realization of the quantum algebra $\mathcal{U}_q(sl_2)$ is obtained by taking \cite{Biedenharn-1989}
\begin{align}
\label{Realization}
J_{+}=q^{-\frac{A_0+B_0-1}{4}}A_{+}B_{-},\qquad J_{-}=q^{-\frac{A_0+B_0-1}{4}}A_{-}B_{+},\qquad J_0=\frac{A_0-B_0}{2}.
\end{align}
It can be verified, using the commutation relations \eqref{Algebra}, that the generators \eqref{Realization} satisfy the defining relations of $U_{q}(sl_{2})$ which read
\begin{align*}
[J_0,J_{\pm}]=\pm J_{\pm},\qquad  [J_{+},J_{-}]=\frac{q^{J_0}-q^{-J_0}}{q^{1/2}-q^{-1/2}}.
\end{align*}
Upon taking $k=q^{2J_0}$, $e=J_{+}$ and $f=J_{-}$, the Chevalley presentation is obtained \cite{Vilenkin-1991}:
\begin{align*}
kk^{-1}=k^{-1}k=1,\quad k^{1/2}e=qek^{1/2},\quad k^{1/2}f=q^{-1}fk^{1/2},\quad [e,f]=\frac{k^{1/2}-k^{-1/2}}{q^{1/2}-q^{-1/2}}.
\end{align*}
The representation of the oscillator algebra \eqref{Rep} on the states \eqref{States} can be used to construct a representation of $\mathcal{U}_q(sl_{2})$. Let $N$ be a non-negative integer and consider the $(N+1)$-dimensional vector space spanned by the states
\begin{align}
\label{Rep-Space}
\kket{n}{N}\equiv\ket{n,N-n},\qquad n=0,\ldots,N.
\end{align}
The states \eqref{Rep-Space} are orthonormal, i.e.
\begin{align*}
\bbraket{N'}{n'}{n}{N}=\delta_{nn'}\delta_{NN'}.
\end{align*}
It follows from \eqref{Rep} that the action of the $\mathcal{U}_q(sl_{2})$ generators \eqref{Realization} on the basis vectors \eqref{Rep-Space} is given by
\begin{align}
\label{UqRep}
\begin{aligned}
J_+\kket{n}{N}&=q^{(1-N)/4}\sqrt{\frac{(1-q^{n+1})}{1-q}\frac{(1-q^{N-n})}{1-q}}\;\kket{n+1}{N},
\\
J_-\kket{n}{N}&=q^{(1-N)/4}\sqrt{\frac{(1-q^{n})}{1-q}\frac{(1-q^{N-n+1})}{1-q}}\;\kket{n-1}{N},
\\
J_0\kket{n}{N}&=(n-N/2)\kket{n}{N}.
\end{aligned}
\end{align}
The actions \eqref{UqRep} correspond to the finite-dimensional irreducible representations of $\mathcal{U}_q(sl_2)$ \cite{Vilenkin-1991}. Hence the direct product states \eqref{Rep-Space} of two independent $q$-oscillators with fixed sums of the quantum numbers $n_{A}$, $n_{B}$ support the irreducible representations of the quantum algebra $\mathcal{U}_q(sl_2)$. 
\subsection{Unitary $q$-rotation operators and matrix elements}
Let us now construct, as in \cite{Zhedanov-1993-06}, the unitary $q$-rotation operators. In analogy with Lie theory, we seek to construct these operators as $q$-exponentials in the generators. Upon using the conjugation formula \eqref{Conjugation-2} and the commutation relations \eqref{Algebra}, a straightforward calculation shows that
\begin{align*}
e_{q}(\alpha\,A_{-}B_{+})\,A_{+}B_{-}\,E_{q}(-\alpha\,A_{-}B_{+})=A_{+}B_{-}+\frac{\alpha}{(1-q)^2}\,q^{A_0}\,-\frac{\alpha}{(1-q)^2}\frac{1}{1-\alpha\,A_-B_{+}}q^{B_0},
\end{align*}
and
\begin{align*}
e_{q}(\alpha\,A_{-}B_{+})\,q^{B_0}\,E_{q}(-\alpha\,A_{-}B_{+})=\frac{1}{1-\alpha\,A_{-}B_{+}}q^{B_0},
\end{align*}
where the formal substitution $\sum_{n} X^{n}=\frac{1}{1-X}$ was made. Combining the above identities, one finds
\begin{align*}
e_{q}(\alpha\,A_{-}B_{+})\left[A_{+}B_{-}+\frac{\alpha}{(1-q)^2}q^{B_0}\right]E_{q}\left(-\alpha\,A_{-}B_{+}\right)=A_{+}B_{-}+\frac{\alpha}{(1-q)^2}q^{A_0}.
\end{align*}
and hence one has
\begin{align*}
e_{q}(\alpha\,A_{-}B_{+})\,e_{q}\left(\beta A_{+}B_{-}+\frac{\alpha \beta}{(1-q)^2}q^{B_0}\right)=e_{q}\left(\beta A_{+}B_{-}+\frac{\alpha \beta}{(1-q)^2}q^{A_0}\right)\,e_{q}(\alpha\,A_{-}B_{+}).
\end{align*}
Since 
\begin{align*}
A_{+}B_{-}q^{B_0}=q q^{B_0}A_{+}B_{-},\qquad \text{and}\qquad q^{A_0}A_{+}B_{-}=q A_{+}B_{-}q^{A_0},
\end{align*}
it follows from the identities \eqref{Q-Commutation} that
\begin{align}
\label{Master-1}
e_{q}(\alpha\,A_{-}B_{+})\,e_{q}\left(\frac{\alpha\beta}{(1-q)^2}q^{B_0}\right)\,e_{q}(\beta A_{+}B_{-})=e_{q}(\beta A_{+}B_{-})e_{q}\left(\frac{\alpha\beta}{(1-q)^2}q^{A_0}\right)\,e_{q}(\alpha A_{-}B_{+}).
\end{align}
Inverting the relation \eqref{Master-1}, one finds a similar relation involving big $q$-exponentials
\begin{align}
\label{Master-2}
E_{q}(\gamma A_{+}B_{-})E_{q}\left(-\frac{\gamma\delta}{(1-q)^2}q^{B_0}\right)E_{q}(\delta A_{-}B_{+})=E_{q}(\delta A_{-}B_{+})E_{q}\left(-\frac{\gamma \delta}{(1-q)^2}q^{A_0}\right)E_{q}(\gamma A_{+}B_{-}).
\end{align}
Let $\theta$ be a real number such that $|\theta|<1$ and consider the unitary operator
\begin{align}
\label{Main-Operator}
U(\theta)=e_{q}^{1/2}(\theta^2 q^{B_0})\,e_{q}(\theta (1-q)A_{+}B_{-})\,E_{q}(-\theta(1-q)A_{-}B_{+})\,E_{q}^{1/2}(-\theta^2 q^{A_0}).
\end{align}
The relation $U^{\dagger}U=1$ follows from \eqref{Master-1} and the relation $UU^{\dagger}=1$ follows from \eqref{Master-2}. Acting by conjugation on the generators \eqref{Realization} in the Schwinger realization, the operator \eqref{Main-Operator} generates automorphisms of $\mathcal{U}_q(sl_2)$ (see \cite{Zhedanov-1993-06} for details). In light of the $2:1$ homomorphism between $SU(2)$ and $SO(3)$, the unitary operator \eqref{Main-Operator} will be referred to as a ``$q$-rotation''' as it is a $q$-extension of a $SU(2)$ element obtained via the exponential map from the algebra to the group. Indeed, upon using the relations
\begin{align*}
\lim_{q\rightarrow 1}\frac{e_{q}(\theta^2 q^{B_0})}{e_q(\theta^2)}=(1-\theta^2)^{B_0}=\exp\left(\log\, (1-\theta^2)\,\widetilde{B}_0\right),
\end{align*}
and
\begin{align*}
\lim_{q\rightarrow 1}\frac{E_{q}(-\theta^2 q^{A_0})}{E_q(-\theta^2)}=(1-\theta^2)^{-A_0}=\exp\left(-\log\,(1-\theta^2)\,\widetilde{A}_0\right),
\end{align*}
as well as the standard Baker-Campbell-Hausdorff relation \cite{Gilmore-2006}, the limit as $q\uparrow 1$ of the unitary operator \eqref{Main-Operator} is seen to be
\begin{align}
\label{Limit-Operator}
\lim_{q\rightarrow 1} U(\theta)=\exp\left(\frac{\theta}{\sqrt{1-\theta^2}}\,\widetilde{A}_{+}\,\widetilde{B}_{-}\right)\exp\left(-\log\,(1-\theta^2)\,\frac{\widetilde{A}_0-\widetilde{B}_0}{2}\right)\exp\left(-\frac{\theta}{\sqrt{1-\theta^2}}\,\widetilde{A}_{-}\widetilde{B}_{+}\right),
\end{align}
where $\widetilde{A}_{\pm}$ and $\widetilde{B}_{\pm}$ satisfy the standard oscillator commutation relations \cite{Cohen-1973}. Since the operators
\begin{align*}
\widetilde{J}_0=\frac{\widetilde{A}_0-\widetilde{B}_0}{2},\qquad \widetilde{J}_+=\widetilde{A}_+\widetilde{B}_{-},\qquad \widetilde{J}_{-}=\widetilde{A}_{-}\widetilde{B}_{+},
\end{align*}
satisfy the $\mathfrak{su}(2)$ commutation relations
\begin{align*}
[\widetilde{J}_0,\widetilde{J}_{\pm}]=\pm \widetilde{J}{\pm},\qquad [\widetilde{J}_{+},\widetilde{J}_{-}]=2\widetilde{J}_0,
\end{align*}
one finds that upon taking $\theta=\sin\tau$, the operator \eqref{Limit-Operator} has the expression
\begin{align*}
\lim_{q\rightarrow 1} U(\sin \tau)=\exp(\tan \tau \widetilde{J}_{+})\exp(-2\,\log(\cos \tau) \widetilde{J}_0)\exp(-\tan \tau \widetilde{J}_-).
\end{align*}
From the disentangling formulas for $SU(2)$ \cite{Truax-1985}, one finally obtains
\begin{align*}
\lim_{q\rightarrow 1} U(\sin \tau)=\exp(\tau\,(\widetilde{J}_{+}-\widetilde{J}_-)),
\end{align*}
which corresponds to a $SU(2)$ group element.

In the following we will focus on the matrix elements of the unitary operator $U(\theta)$ given in \eqref{Main-Operator} in the basis \eqref{Rep-Space} of irreducible representations of $\mathcal{U}_q(sl_2)$; these matrix elements will be denoted by
\begin{align}
\label{Matrix-Elements}
\chi_{n,x}^{(N)}=\BBraket{N}{n}{U(\theta)}{x}{N},
\end{align}
where $n,x\in \{0,1,\ldots,N\}$.
\section{Matrix elements and self-duality}
In this section, the matrix elements \eqref{Matrix-Elements} of the $q$-rotation operators \eqref{Main-Operator} are obtained by a direct calculation and are shown to involve the quantum  $q$-Krawtchouk polynomials. The weight function and the orthogonality relation satisfied by these polynomials are derived from the properties of $U(\theta)$. A self-duality relation for the matrix elements is also obtained and the $q\uparrow 1$ limit is examined.
\subsection{Matrix elements and quantum $q$-Krawtchouk polynomials}
To obtain the explicit expression of the matrix elements \eqref{Matrix-Elements} one can proceed directly by expanding the $q$-exponentials in \eqref{Main-Operator} according to \eqref{Little-Exp}, \eqref{Big-Exp}  and use the actions \eqref{Rep} of the generators on the basis vectors \eqref{Rep-Space}. To perform this calculation, it is useful to note that
\begin{subequations}
\label{Actions}
\begin{align}
\label{Action-A}
(A_{-}B_{+})^{\alpha}\kket{x}{N}=(1-q)^{-\alpha}\sqrt{\frac{(q;q)_{x}}{(q;q)_{x-\alpha}}\frac{(q;q)_{N-x+\alpha}}{(q;q)_{N-x}}}\;\kket{x-\alpha}{N},
\end{align}
and
\begin{align}
\label{Action-B}
(A_{+}B_{-})^{\beta}\kket{x}{N}=(1-q)^{-\beta}\sqrt{\frac{(q;q)_{x+\beta}}{(q;q)_{x}}\frac{(q;q)_{N-x}}{(q;q)_{N-x-\beta}}}\;\kket{x+\beta}{N}.
\end{align}
\end{subequations}
Upon expanding the operator \eqref{Main-Operator} according to \eqref{Little-Exp} and \eqref{Big-Exp}, using the actions \eqref{Actions}, reversing the order of the first summation and exchanging the summation order, one finds
\begin{multline*}
U(\theta)\kket{x}{N}=\sum_{n=0}^{N}\left\{E_{q}(-\theta^2 q^{x})\;e_{q}(\theta^2 q^{N-n})\;\frac{(q;q)_{x}(q;q)_{n}}{(q;q)_{N-x}(q;q)_{N-n}}\right\}^{1/2}(-1)^{x}(\theta)^{x+n}
\\
\times \sum_{\gamma=0}^{x}\frac{(-1/\theta^2)^{\gamma}}{(q;q)_{\gamma}}\,q^{\binom{x-\gamma}{2}}\,\frac{(q;q)_{N-\gamma}}{(q;q)_{x-\gamma}(q;q)_{n-\gamma}}\,\kket{n}{N}.
\end{multline*}
Upon using the identity \eqref{Identity-1} for the $q$-shifted factorials, the formulas \eqref{Little-Exp}, \eqref{Big-Exp} and \eqref{QBinom} for the $q$-exponentials and the $q$-binomial coefficients as well as the definition \eqref{Basic-Hyper} for the basic hypergeometric series, one finds from the above the following expression for the matrix elements:
\begin{align}
\label{Formula-1}
\chi_{n,x}^{(N)}=(-1)^{x}\theta^{n+x}q^{\binom{x}{2}}\qbinom{N}{x}^{1/2}\qbinom{N}{n}^{1/2}\frac{(\theta^2;q)_{N-n}^{1/2}}{(\theta^2;q)_{x}^{1/2}}\;\;\pfq{2}{1}{q^{-n},q^{-x}}{q^{-N}}{q,\,\frac{q^{n+1}}{\theta^2q^{N}}}.
\end{align}
The quantum $q$-Krawtchouk $K_{n}^{\text{Qtm}}(q^{-x};p,N;q)$ of degree $n$ in the variable $q^{-x}$ are defined by \cite{Koekoek-2010}
\begin{align}
\label{Aff-Def}
K_{n}^{\text{Qtm}}(q^{-x};p,N;q)=\pfq{2}{1}{q^{-n},q^{-x}}{q^{-N}}{q,\,p\,q^{n+1}}.
\end{align}
Comparing the definition \eqref{Aff-Def} with the formula \eqref{Formula-1}, it follows that the matrix elements \eqref{Matrix-Elements} of the unitary $q$-rotation operator \eqref{Main-Operator} can be written as
\begin{align}
\label{Matrix-Elements-Explicit}
\chi_{n,x}^{(N)}=(-1)^{x}\theta^{n+x}q^{\binom{x}{2}}\qbinom{N}{n}^{1/2}\qbinom{N}{x}^{1/2}\frac{(\theta^2;q)_{N-n}^{1/2}}{(\theta^2;q)_{x}^{1/2}}\;\;K_{n}^{\text{Qtm}}\left(q^{-x};\frac{1}{\theta^2 q^{N}},N;q\right).
\end{align}
This result can be compared with those of \cite{Miller-1994}. The matrix elements can be cast in the form
\begin{align}
\label{Canonical-Form}
\chi_{n,x}^{(N)}=(-1)^{x}\sqrt{w_{x}^{(N)}}\;\;\widehat{K}_{n}^{\text{Qtm}}\left(q^{-x};\frac{1}{\theta^2 q^{N}},N;q\right),
\end{align}
where $w_{x}^{(N)}$ is a $q$-analog of the binomial distribution
\begin{align}
\label{Weight}
w_{x}^{(N)}=\left[\chi_{0,x}^{(N)}\right]^2=\qbinom{N}{x}\frac{(\theta^2;q)_{N}}{(\theta^2;q)_{x}}\theta^{2x}q^{x(x-1)},
\end{align}
and where $\widehat{K}_n^{\text{Qtm}}\left(q^{-x};\frac{1}{\theta^2 q^{N}},N;q\right)$ are the normalized quantum $q$-Krawtchouk polynomials
\begin{align}
\label{Normalized-QTM}
\widehat{K}_n^{\text{Qtm}}\left(q^{x};\frac{1}{\theta^2 q^{N}},N;q\right)=\sqrt{\qbinom{N}{n}\frac{\theta^{2n}}{(\theta^2 q^{N-n};q)_{n}}}\;\;K_{n}^{\text{Qtm}}\left(q^{-x};\frac{1}{\theta^2 q^{N}}, N;q\right).
\end{align}
The orthonormality of the basis states \eqref{Rep-Space} and the unitarity of the operator \eqref{Main-Operator} directly lead to a pair of orthogonality relations for the quantum  $q$-Krawtchouk polynomials. In fact, one has
\begin{align*}
\BBraket{N}{n'}{UU^{\dagger}}{n}{N}&=\sum_{x=0}^{N}\BBraket{N}{n}{U}{x}{N}\;\BBraket{N}{x}{U^{\dagger}}{n}{N}=\sum_{k=0}^{N}\chi_{n,x}^{(N)}\;[\chi_{n',x}^{(N)}]^{*}=\delta_{nn'},
\\
\BBraket{N}{x'}{U^{\dagger}U}{x}{N}&=\sum_{n=0}^{N}\BBraket{N}{x'}{U^{\dagger}}{n}{N}\BBraket{N}{n}{U}{x}{N}=\sum_{n=0}^{N}\chi_{n,x}^{(N)}[\chi_{n,x'}^{(N)}]^{*}=\delta_{xx'},
\end{align*}
where $z^{*}$ stands for complex conjugation. Since the matrix elements are real, it follows from the above that the quantum  $q$-Krawtchouk polynomials \eqref{Normalized-QTM} satisfy the orthogonality relation
\begin{align}
\label{Ortho-1}
\sum_{x=0}^{N}w_{x}^{(N)}\;K_n^{\text{Qtm}}\left(q^{-x};\frac{1}{\theta^2q^{N}},N;q\right)\,K_{n'}^{\text{Qtm}}\left(q^{-x};\frac{1}{\theta^2 q^{N}},N;q\right)=\delta_{nn'}\frac{(q;q)_{n}(q;q)_{N-n}(\theta^2q^{N-n};q)_{n}}{(q;q)_{N}\;\theta^{2n}},
\end{align}
with respect to the weight function \eqref{Weight} and the dual orthogonality relation
\begin{align}
\label{Ortho-2}
\sum_{n=0}^{N}\qbinom{N}{n}\frac{\theta^{2n}}{(\theta^2q^{N-n};q)_{n}}\;K_n^{\text{Qtm}}\left(q^{-x};\frac{1}{\theta^2q^{N}},N;q\right)\,K_{n}^{\text{Qtm}}\left(q^{-x'};\frac{1}{\theta^2 q^{N}},N;q\right)=\frac{\delta_{xx'}}{\omega_{x}^{(N)}}.
\end{align}
\subsection{Duality}
The matrix elements \eqref{Matrix-Elements} have a self-duality property which can be obtained as follows. Using the reality of the matrix elements \eqref{Matrix-Elements} and the unitary of the $q$-rotation operator \eqref{Main-Operator}, one can write
\begin{align*}
\chi_{n,x}^{(N)}=\left[\BBraket{N}{n}{U}{x}{N}\right]^{*}=\BBraket{N}{x}{U^{\dagger}}{n}{N}=\Braket{x,N-x}{\;U^{-1}\;}{n,N-n},
\end{align*}
where the direct product notation \eqref{States} was used for the last equality. It is easily seen from \eqref{Main-Operator} that the inverse operator $U^{-1}$ is obtained from $U$ by permuting the algebra generators $\{A_{\pm},A_0\}$ with $\{B_{\pm},B_{0}\}$. In view of the definition \eqref{Rep-Space} for the states $\kket{n}{N}$, this observation leads to the duality relation
\begin{align}
\label{Duality}
\chi_{n,x}^{(N)}=\chi_{N-x,N-n}^{(N)}.
\end{align} 
The relation \eqref{Duality} allows to exchange the roles of the variable $x$ and the degree $n$.
\subsection{The $q\uparrow 1$ limit}
The $q\uparrow 1$ limit can be taken in a straightforward fashion in the matrix elements \eqref{Formula-1} using the formula \eqref{Basic-Hyper} for the basic hypergeometric series. With $\theta=\sin \tau$, one finds
\begin{align*}
\lim_{q\rightarrow 1}\chi_{n,x}^{(N)}&=\binom{N}{x}^{1/2}\binom{N}{n}^{1/2}(-1)^{x}\tan^{n+x}\tau \cos^{N}\tau \;\;\pGq{2}{1}{-n,-x}{-N}{\frac{1}{\sin^2\tau}}
\\
&=\binom{N}{n}^{1/2}\binom{N}{x}^{1/2}(-1)^{x}\tan^{n+x}\tau \cos^{N}\tau \;K_{n}(x; \sin^2\tau; N),
\end{align*}
where $K_{n}(x;p;N)$ are the standard Krawtchouk polynomials \cite{Koekoek-2010}.
\section{Structure relations}
In this section, it is shown how the algebraic setting can be used to derive structure relations for the quantum $q$-Krawtchouk polynomials.
\subsection{Backward relation}
Consider the matrix element $\BBraket{N-1}{n}{A_{-}\,U}{x}{N}$. The action \eqref{Rep} gives
\begin{align}
\label{Raising-RHS}
\BBraket{N-1}{n}{A_{-}U}{x}{N}=\sqrt{\frac{1-q^{n+1}}{1-q}}\,\chi_{n+1,x}^{(N)}.
\end{align}
Using the unitarity of $U$, one has also
\begin{align}
\label{Raising-LHS}
\BBraket{N-1}{n}{A_{-}U}{x}{N}=\BBraket{N-1}{n}{U\;U^{\dagger}\,A_{-}U}{x}{N}.
\end{align}
To obtain a backward relation, one needs to calculate $U^{\dagger}A_{-}U$. Making use of \eqref{Master-1}, one has
\begin{multline*}
U^{\dagger}A_{-}U=e_q^{1/2}(\theta^2 q^{A_0})e_{q}(\theta(1-q)A_{-}B_{+}) E_{q}(-\theta(1-q)A_{+}B_{-})
\\
\times A_{-}\; e_{q}(\theta(1-q)A_{+}B_{-})E_{q}(-\theta(1-q)A_{-}B_{+})E_{q}^{1/2}(-\theta^2 q^{A_0}).
\end{multline*}
With the help of formula \eqref{Conjugation-1}, one easily finds
\begin{align*}
E_{q}(-\theta(1-q)A_{+}B_{-})\,A_{-}\,e_{q}(\theta(1-q)A_{+}B_{-})=A_{-}+\theta q^{A_0}B_{-},
\end{align*}
and thus
\begin{align*}
U^{\dagger}A_{-}U=e_q^{1/2}(\theta^2 q^{A_0})e_{q}(\theta(1-q)A_{-}B_{+})\left[A_{-}+\theta q^{A_0}B_{-}\right]E_{q}(-\theta(1-q)A_{-}B_{+})E_{q}^{1/2}(-\theta^2 q^{A_0}).
\end{align*}
The conjugation formula \eqref{Conjugation-2} gives
\begin{align*}
e_{q}(\theta(1-q)A_{-}B_{+})\,q^{A_0}B_{-}\,E_{q}(-\theta(1-q)A_{-}B_{+})&=q^{A_0}B_{-}-\theta q^{A_0}A_{-},
\end{align*}
and consequently
\begin{align*}
U^{\dagger}A_{-}U=e_{q}^{1/2}(\theta^2 q^{A_0})\left[(1-\theta^2 q^{A_0})A_{-}+\theta q^{A_0}B_{-}\right]E_{q}^{1/2}(-\theta^2 q^{A_0}).
\end{align*}
Formally, one has
\begin{align*}
e_q^{1/2}(\theta^2 q^{A_0})\,A_{-}\,E_{q}^{1/2}(-\theta^2 q^{A_0})=\sqrt{\frac{1}{1-\theta^2 q^{A_0}}}A_{-},
\end{align*}
and thus one finally obtains
\begin{align}
\label{Conjugation-Am}
U^{\dagger}\,A_{-}\,U=\sqrt{1-\theta^2 q^{A_0}}\,A_{-}+\theta q^{A_0}B_{-}.
\end{align}
Upon inserting the result \eqref{Conjugation-Am} in \eqref{Raising-LHS} and using the actions \eqref{Rep}, one finds
\begin{align*}
\BBraket{N-1}{n}{A_{-}U}{x}{N}=\sqrt{\frac{(1-q^{x})(1-\theta^2 q^{x-1})}{1-q}}\;\chi_{n,x-1}^{(N-1)}+\theta q^{x}\sqrt{\frac{1-q^{N-x}}{1-q}}\;\chi_{n,x}^{(N-1)}.
\end{align*}
Combining the above relation with \eqref{Raising-RHS}, one obtains the backward relation
\begin{align}
\label{Raising-Relation-n}
\sqrt{1-q^{n+1}}\;\chi_{n+1,x}^{(N)}=\sqrt{(1-q^{x})(1-\theta^2 q^{x-1})}\;\chi_{n,x-1}^{(N-1)}+\theta\,q^{x}\sqrt{1-q^{N-x}}\;\chi_{n,x}^{(N-1)}.
\end{align}
Using the expression \eqref{Canonical-Form} and the formula \eqref{Weight} for the weight function, the relation \eqref{Raising-Relation-n} gives for the quantum $q$-Krawtchouk polynomials 
\begin{multline}
\label{Raising-Relation-n-Polynomials}
(1-q^{N})\,K_{n+1}^{\text{Qtm}}\left(q^{-x};\frac{1}{\theta^2 q^{N}},N;q\right)=(q^{x}-q^{N})\,K_{n}^{\text{Qtm}}\left(q^{-x};\frac{1}{\theta^2 q^{N-1}},N-1;q\right)\\
+\frac{q}{\theta^2}(1-q^{-x})(1-\theta^2 q^{x-1})\,K_{n}^{\text{Qtm}}\left(q^{-(x-1)};\frac{1}{\theta^2 q^{N-1}},N-1;q\right).
\end{multline}
The backward relation \eqref{Raising-Relation-n-Polynomials} can be used to generate polynomials recursively and coincides with the one given in \cite{Koekoek-2010}.
\subsection{Forward relation}
Consider the matrix element $\BBraket{N}{n}{A_{+}U}{x}{N-1}$. The action \eqref{Rep} gives
\begin{align}
\label{Lowering-RHS}
\BBraket{N}{n}{A_{+}U}{x}{N-1}=\sqrt{\frac{1-q^{n}}{1-q}}\;\chi_{n-1,x}^{(N-1)}.
\end{align}
Taking the conjugate of \eqref{Conjugation-Am}, one has
\begin{align*}
U^{\dagger}\,A_{+}\,U=A_{+}\sqrt{1-\theta^2 q^{A_0}}+\theta q^{A_0}B_{+},
\end{align*}
which upon using \eqref{Rep} yields
\begin{align}
\label{Lowering-LHS}
\BBraket{N}{n}{U\;U^{\dagger}A_{+}U}{x}{N-1}=\sqrt{\frac{(1-q^{x+1})(1-\theta^2 q^{x})}{1-q}}\;\chi_{n,x+1}^{(N)}+\theta\,q^{x}\sqrt{\frac{1-q^{N-x}}{1-q}}\,\chi_{n,x}^{(N)}.
\end{align}
Comparing \eqref{Lowering-LHS} with \eqref{Lowering-RHS}, one obtains the forward relation
\begin{align}
\label{Lowering-Relation-n}
\sqrt{1-q^{n}}\;\chi_{n-1,x}^{(N-1)}=\sqrt{(1-q^{x+1})(1-\theta^2q^{x})}\;\chi_{n,x+1}^{(N)}+\theta\,q^{x}\sqrt{1-q^{N-x}}\;\chi_{n,x}^{(N)}.
\end{align}
For the quantum $q$-Krawtchouk polynomials, the relation \eqref{Lowering-Relation-n} translates into
\begin{multline}
\label{Lowering-Relation-n-Polynomials}
\left(\frac{1-q^{n}}{1-q^{N}}\right)\,K_{n-1}^{\text{Qtm}}\left(q^{-x};\frac{1}{\theta^2 q^{N-1}},N-1;q\right)=
\\
\theta^2\,q^{x}\,K_{n}^{\text{Qtm}}\left(q^{-x};\frac{1}{\theta^2 q^{N}},N;q\right)-\theta^2 q^{x}\,K_{n}^{\text{Qtm}}\left(q^{-(x+1)};\frac{1}{\theta^2 q^{N}},N;q\right).
\end{multline}
It is verified that \eqref{Lowering-Relation-n-Polynomials} corresponds to the one found in \cite{Koekoek-2010}.
\subsection{Dual backward and forward relations}
The self-duality property \eqref{Duality} can be exploited to derive additional relations from the backward and forward relations \eqref{Raising-Relation-n} and \eqref{Lowering-Relation-n} satisfied by the matrix elements. From \eqref{Duality}, one finds that 
\begin{align*}
\sqrt{1-q^{N-x}}\;\chi_{n,x}^{(N)}&=\sqrt{(1-q^{N-n})(1-\theta^2 q^{N-n-1})}\;\chi_{n,x}^{(N-1)}+\theta q^{N-n}\sqrt{1-q^{n}}\;\chi_{n-1,x}^{(N-1)},
\\
\sqrt{1-q^{N-x}}\;\chi_{n-1,x}^{(N-1)}&=\sqrt{(1-q^{N-n+1})(1-\theta^2 q^{N-n})}\;\chi_{n-1,x}^{(N)}+\theta q^{N-n}\sqrt{1-q^{n}} \;\chi_{n,x}^{(N)},
\end{align*}
which translate into other type of identities for the quantum $q$-Krawtchouk polynomials. Equivalently, one can consider matrix elements of the form $\BBraket{N}{n}{U\,B_{\pm}}{x}{N}$ and use the identities
\begin{align}
\label{Conjug-Iden-2}
UB_{+}U^{\dagger}=B_{+}\sqrt{1-\theta^2 q^{B_0}} +\theta q^{B_0}A_{+},\quad \text{and}\quad UB_{-}U^{\dagger}=\sqrt{1-\theta^2 q^{B_0}}B_{-} +\theta q^{B_0}A_{-}.
\end{align}
\section{Generating function}
In this section, two generating functions for the quantum $q$-Krawtchouk are derived. The first one generates the polynomials with respect to the degrees and the other with respect to the variables.
\subsection{Generating function with respect to the degrees}
Consider the matrix element $\BBraket{N}{0}{V(t)\,U(\theta)}{x}{N}$ where $V(t)$ is the operator
\begin{align*}
V(t)=E_{q}(t(1-q)A_{-}B_{+})E_{q}^{1/2}(-\theta^2 q^{B_0}).
\end{align*}
Upon expanding the big $q$-exponentials according to \eqref{Big-Exp}, using the action \eqref{Action-A} and the definition \eqref{Matrix-Elements} of the matrix elements of $U(\theta)$, it is directly checked that
\begin{align*}
\BBraket{N}{0}{V(t)\,U(\theta)}{x}{N}=
\sum_{n=0}^{N}\qbinom{N}{n}^{1/2}\,E_{q}^{1/2}(-\theta^2 q^{N-n})\;q^{n(n-1)/2}\;\chi_{n,x}^{(N)}\;t^{n}.
\end{align*}
With the identity
\begin{align*}
E_{q}(-\lambda\;q^{n})=\frac{E_{q}(-\lambda)}{(\lambda;q)_{n}},
\end{align*}
and the explicit expression \eqref{Matrix-Elements-Explicit} for the matrix elements $\chi_{n,x}^{(N)}$, one can write
\begin{multline}
\label{Gen-1-RHS}
\BBraket{N}{0}{V(t)U(\theta)}{x}{N}=
\\
\frac{(-\theta)^{x}q^{x(x-1)/2}}{(\theta^2;q)_{x}^{1/2}}\,\qbinom{N}{x}^{1/2}\,E_{q}^{1/2}(-\theta^2)\; \sum_{n=0}^{N}\qbinom{N}{n}q^{n(n-1)/2}\,K_{n}^{\text{Qtm}}\left(q^{-x};\frac{1}{\theta^2 q^{N}},N; q\right)\;(\theta\, t)^{n},
\end{multline}
which has the form of a generating function for the quantum $q$-Krawtchouk polynomials. Let us compute the matrix element $\BBraket{N}{0}{V(t)\,U(\theta)}{x}{N}$ in a different way. It follows from \eqref{Master-2} that
\begin{align}
\label{Iden-1}
E_{q}(\gamma A_{-}B_{+})e_{q}(-\delta A_{+}B_{-})=e_{q}\left(\frac{\gamma \delta}{(1-q)^2}q^{B_0}\right)\,e_{q}(-\delta A_{+}B_{-})\,E_{q}(\gamma A_{-}B_{+})\,E_{q}\left(\frac{-\gamma \delta}{(1-q)^2}q^{A_0}\right),
\end{align}
With $\gamma=t(1-q)$ and $\delta=-\theta(1-q)$, the above identity gives
\begin{multline*}
V(t)U(\theta)=
\\
e_{q}(-\theta\,t q^{B_0})e_{q}(\theta(1-q)A_{+}B_{-})E_{q}(t(1-q) A_{-}B_{+})E_{q}(\theta\,t q^{A_0})E_{q}(-\theta(1-q)A_{-}B_{+})E_{q}^{1/2}(-\theta^2 q^{A_0}),
\end{multline*}
which leads to the expression
\begin{multline*}
\BBraket{N}{0}{V(t)U(\theta)}{x}{N}=e_q(-\theta\,t q^{N})\,E_{q}^{1/2}(-\theta^2 q^{x})
\\
\BBraket{N}{0}{E_{q}(t(1-q)A_{-}B_{+})\,E_{q}(\theta t q^{A_0})\,E_{q}(-\theta (1-q) A_{-}B_{+})}{x}{N}.
\end{multline*}
Upon using the identity
\begin{align*}
e_{q}(\lambda q^{n})=e_{q}(\lambda)\,(\lambda;q)_{n},
\end{align*}
and the orthonormality of the states, one easily obtains
\begin{multline}
\label{Gen-1-LHS}
\BBraket{N}{0}{V(t)U(\theta)}{x}{N}=
\\
\frac{(-\theta)^x q^{x(x-1)/2}}{(\theta^2;q)_{x}^{1/2}}\qbinom{N}{x}^{1/2}\,E_{q}^{1/2}(-\theta^2)\;(-\theta t;q)_{N}\sum_{\gamma=0}^{x}\frac{(t/\theta)^{\gamma}q^{\gamma(\gamma+1)/2}}{(q;q)_{\gamma}}\,\frac{(q^{-x};q)_{\gamma}}{(-\theta t;q)_{\gamma}}.
\end{multline}
Comparing \eqref{Gen-1-RHS} with \eqref{Gen-1-LHS}, using \eqref{Basic-Hyper} and taking $z=\theta\,t$, one finds the following generating function for the quantum $q$-Krawtchouk polynomials
\begin{align}
\label{Gen-Fun-1}
(-z;q)_{N}\;\pfq{1}{1}{q^{-x}}{-z}{q,\,-\frac{qz}{\theta^2}}=\sum_{n=0}^{N}\qbinom{N}{n}q^{n(n-1)/2}\,K_{n}^{(\text{Qtm})}\left(q^{-x};\frac{1}{\theta^2 q^{N}}, N;q\right)\;z^{n}.
\end{align}
Using the identity
\begin{align}
\label{Fin-1}
(q^{-N};q)_{n}=\frac{(q;q)_{N}}{(q;q)_{N-n}}(-1)^{n}q^{\binom{n}{2}-Nn},
\end{align}
defining $v=-q^{N}z$ and taking $p=\frac{1}{\theta^2 q^{N}}$, the relation \eqref{Gen-Fun-1} takes the form
\begin{align}
\label{Gen-Fun-1-B}
(v q^{-N};q)_{N}\;\pfq{1}{1}{q^{-x}}{vq^{-N}}{q,\;p\,q\,v}=\sum_{n=0}^{N}\frac{(q^{-N};q)_{n}}{(q;q)_{n}}\; K_{n}^{\text{Qtm}}\left(q^{-x};p,N;q\right)\;v^{n}.
\end{align}
The RHS of \eqref{Gen-Fun-1-B} corresponds to one of the generating functions given in  \cite{Koekoek-2010}. In the latter reference however, the LHS is given in terms of a ${}_2\phi_1$ basic hypergeometric series. The results of \cite{Koekoek-2010} can be recovered as follows. Consider the identity (\cite{Gasper-2004} Appendix III):
\begin{align*}
\pfq{2}{1}{q^{-n},b}{c}{q,z}=\frac{(c/b;q)_{n}}{(c;q)_{n}}\,\pfq{3}{2}{q^{-n},bzq^{-n}/c}{b q^{1-n}/c,0}{q,q}.
\end{align*}
With $b\rightarrow\lambda b$ and $z\rightarrow z/\lambda$, taking the limit as $\lambda\rightarrow \infty$ gives the transformation formula
\begin{align*}
\pfq{1}{1}{q^{-n}}{c}{q, t}=\frac{1}{(c;q)_{n}}\,\pfq{2}{1}{q^{-n},tq^{-n}/c}{0}{q,cq^{n}}.
\end{align*}
Upon using the above identity and the relation $\frac{(a q^{-n};q)_{n}}{(a q^{-n};q)_{k}}=(a q^{k-n};q)_{n-k}$, the generating relation \eqref{Gen-Fun-1-B} becomes
\begin{align*}
(v q^{x-N};q)_{N-x}\;\pfq{2}{1}{q^{-x},p q^{N-x+1}}{0}{q,v q^{x-N}}=\sum_{n=0}^{N}\frac{(q^{-N};q)_{n}}{(q;q)_{n}}\; K_{n}^{\text{Qtm}}\left(q^{-x};p,N;q\right)\;v^{n},
\end{align*}
which coincides with the generating function given in \cite{Koekoek-2010}.

\subsection{Generating function with respect to the variables}
To obtain a generating function where the sum is performed on the variables, one can consider the matrix element $\BBraket{N}{n}{U(\theta)W(t)}{0}{x}$ where
\begin{align*}
W(t)=e_{q}^{1/2}(\theta^2 q^{A_0})\,e_q(t(1-q)A_{+}B_{-}).
\end{align*}
On the one hand, expanding the $q$-exponentials and using \eqref{Matrix-Elements-Explicit} yields
\begin{align}
\label{Gen-2-RHS}
\BBraket{N}{n}{U(\theta)W(t)}{0}{N}=e_{q}^{1/2}(\theta^2 q^{N-n})\;\theta^{n}\;\qbinom{N}{n}^{1/2}\sum_{x=0}^{N}\qbinom{N}{x} (-\theta t)^{x}q^{x(x-1)/2}\;K_{n}^{\text{Qtm}}\left(q^{-x};\frac{1}{\theta^2 q^{N}},N;q\right),
\end{align}
which has the form of a generating function. On the other hand, the identity  \eqref{Iden-1} gives
\begin{multline*}
\BBraket{N}{n}{U(\theta) W(t)}{x}{N}=
\\
e_{q}^{1/2}(\theta^2 q^{N-n})\;E_{q}(-\theta t)\;\BBraket{N}{n}{e_{q}(\theta (1-q)A_{+}B_{-})\;e_q(\theta tq^{B_0})\;e_{q}(t(1-q)A_{+}B_{-})}{0}{N}.
\end{multline*}
With the identity \eqref{Identity-1}, one directly finds from the above
\begin{align}
\label{Gen-2-LHS}
\BBraket{N}{n}{U(\theta)W(t)}{0}{x}=
e_{q}^{1/2}(\theta^2 q^{N-n})\;\theta^{n}\;\qbinom{N}{n}^{1/2}\;(\theta t;q)_{N}\;\sum_{\gamma=0}^{N}\frac{\left(\frac{q^{n-N+1}}{\theta^2}\right)^{\gamma}}{(q;q)_{\gamma}}\frac{(q^{-n};q)_{\gamma}}{(\frac{q^{1-N}}{\theta t};q)_{\gamma}}.
\end{align}
Upon comparing \eqref{Gen-2-LHS} with \eqref{Gen-2-RHS} and taking $z=-\theta t$, one obtains the generating relation
\begin{align}
\label{Gen-2}
(-z;q)_{N}\;\pfq{2}{1}{q^{-n},0}{-\frac{q^{1-N}}{z}}{q,\,\frac{q^{n+1}}{\theta^2 q^{N}}}=\sum_{x=0}^{N}\qbinom{N}{x}q^{x(x-1)/2}\;K_{n}^{\text{Qtm}}\left(q^{-x};\frac{1}{\theta^2 q^{N}},N;q\right)\;z^{x}.
\end{align}
Using \eqref{Fin-1}, defining $w=-q^{N}z$ and taking $p=\frac{1}{\theta^2 q^{N}}$, one writes \eqref{Gen-2} as
\begin{align}
(w q^{-N};q)_{N}\;\pfq{2}{1}{q^{-n},0}{\frac{q}{w}}{q,\,p\,q^{n+1}}=\sum_{x=0}^{N}\frac{(q^{-N};q)_{x}}{(q;q)_{x}}\,K_{n}^{\text{Qtm}}(q^{-x};p,N;q)\;w^{x}.
\end{align}

\section{Recurrence relation and difference equation}
In this section, the recurrence relation and the difference equation satisfied by the matrix elements $\chi_{n,x}^{(N)}$ of the unitary $q$-rotation operators $U(\theta)$ are obtained and the corresponding relations for the quantum $q$-Krawtchouk polynomials are recovered.
\subsection{Recurrence relation}
To obtain a recurrence relation for the matrix elements $\chi_{n,x}^{(N)}$, one may consider the matrix element $\BBraket{N}{n}{U\,B_{+}B_{-}}{x}{N}$. On the one hand, one has
\begin{align}
\label{Recurrence-RHS}
\BBraket{N}{n}{U\,B_{+}B_{-}}{x}{N}=\left(\frac{1-q^{N-x}}{1-q}\right)\;\chi_{n,x}^{(N)}.
\end{align}
On the other hand, the conjugation identities \eqref{Conjug-Iden-2} give
\begin{multline*}
UB_{+}B_{-}U^{\dagger}=
\\
\Big(B_{+}(1-\theta^2 q^{B_0})B_{-}+\theta B_{+}\sqrt{1-\theta^2 q^{B_0}}q^{B_0}A_{-}+\theta q^{B_0}A_{+}\sqrt{1-\theta^2 q^{B_0}}B_{-}+\theta^2 q^{2B_0}A_{+}A_{-}\Big),
\end{multline*}
and thus
\begin{multline}
\label{Recurrence-LHS}
\BBraket{N}{n}{U\,B_{+}B_{-}}{x}{N}=\left(\frac{(1-q^{N-n})(1-\theta^2 q^{N-n-1})}{1-q}\right)\,\chi_{n,x}^{(N)}
\\
+\theta\;q^{N-n-1}\sqrt{\frac{(1-q^{n+1})(1-q^{N-n})(1-\theta^2 q^{N-n-1})}{(1-q)^2}}\;\chi_{n+1,x}^{(N)}
\\
+\theta q^{N-n}\sqrt{\frac{(1-q^{n})(1-\theta^2 q^{N-n})(1-q^{N-n+1})}{(1-q)^2}}\;\chi_{n-1,x}^{(N)}+\theta^2 q^{2(N-n)}\left(\frac{1-q^{n}}{1-q}\right)\chi_{n,x}^{(N)}.
\end{multline}
Comparing \eqref{Recurrence-LHS} and \eqref{Recurrence-RHS}, one finds that the matrix elements satisfy the recurrence relation
\begin{multline}
\label{Recurrence-Matrix}
(1-q^{N-x})\;\chi_{n,x}^{(N)}=(1-q^{N-n})(1-\theta^2 q^{N-n-1})\;\chi_{n,x}^{(N)}
\\
+\theta q^{N-n-1}\sqrt{(1-q^{n+1})(1-q^{N-n})(1-\theta^2 q^{N-n-1})}\;\chi_{n+1,x}^{(N)}
\\
+\theta q^{N-n}\sqrt{(1-q^{n})(1-\theta^2 q^{N-n})(1-q^{N-n+1})}\;\chi_{n-1,x}^{(N)}+\theta^2 q^{2(N-n)}(1-q^{n})\;\chi_{n,x}^{(N)}.
\end{multline}
Using the expression \eqref{Matrix-Elements-Explicit}, one finds that the recurrence relation for the quantum $q$-Krawtchouk polynomials is of the form
\begin{multline}
\label{Recurrence-Poly-1}
(1-q^{N-x})\;K_{n}^{\text{Qtm}}\left(q^{-x};\frac{1}{\theta^2 q^{N}},N;q\right)=\theta^2 q^{N-n-1}(1-q^{N-n})\;K_{n+1}^{\text{Qtm}}\left(q^{-x};\frac{1}{\theta^2 q^{N}},N;q\right)
\\
+\Big[(1-\theta^2 q^{N-n-1})(1-q^{N-n})-\theta^2q^{2(N-n)}(1-q^{n})\Big]\;K_{n}^{\text{Qtm}}\left(q^{-x};\frac{1}{\theta^2 q^{N}},N;q\right)
\\
+q^{N-n}(1-\theta^2 q^{N-n})(1-q^{n})\;K_{n-1}^{\text{Qtm}}\left(q^{-x};\frac{1}{\theta^2 q^{N}},N;q\right).
\end{multline}
It can be checked that the recurrence relation \eqref{Recurrence-Poly-1} coincides with the one given in \cite{Koekoek-2010}.
\subsection{Difference equation}
To obtain the difference equation satisfied by the matrix elements $\chi_{n,x}^{(N)}$ and consequently by the quantum $q$-Krawtchouk polynomials, one could consider the matrix element
$\BBraket{N}{n}{A_{+}A_{-}U}{x}{N}$ and use the conjugation identities \eqref{Conjugation-1} and \eqref{Conjugation-2} to compute $U^{\dagger}A_{+}A_{-}U$. Alternatively, one can start from the recurrence relation \eqref{Recurrence-Matrix} and use the duality relation \eqref{Duality}. Applying the duality on \eqref{Recurrence-Matrix}, one finds
\begin{multline*}
(1-q^{N-x})\;\chi_{N-x,N-n}^{(N)}=(1-q^{N-n})(1-\theta^2 q^{N-n-1})\;\chi_{N-x,N-n}^{(N)}
\\
+\theta q^{N-n-1}\sqrt{(1-q^{n+1})(1-q^{N-n})(1-\theta^2 q^{N-n-1})}\;\chi_{N-x,N-n-1}^{(N)}
\\
+\theta q^{N-n}\sqrt{(1-q^{n})(1-\theta^2 q^{N-n})(1-q^{N-n+1})}\;\chi_{N-x,N-n+1}^{(N)}+\theta^2 q^{2(N-n)}(1-q^{n})\;\chi_{N-x,N-n}^{(N)}.
\end{multline*}
Upon taking $x\rightarrow N-n$ and $n\rightarrow N-x$, one obtains the following difference equation for the matrix elements $\chi_{n,x}^{(N)}$:
\begin{multline}
\label{Difference-Matrix-Elements}
(1-q^{n})\;\chi_{n,x}^{(N)}=(1-q^{x})(1-\theta^2 q^{x-1})\;\chi_{n,x}^{(N)}
\\
+\theta q^{x-1}\sqrt{(1-q^{N-x+1})(1-q^{x})(1-\theta^2 q^{x-1})}\;\chi_{n,x-1}^{(N)}
\\
+\theta q^{x}\sqrt{(1-q^{N-x})(1-\theta^2 q^{x})(1-q^{x+1})}\;\chi_{n,x+1}^{(N)}+\theta^2 q^{2x}(1-q^{N-x})\;\chi_{n,x}^{(N)}.
\end{multline}
Using the expression \eqref{Matrix-Elements-Explicit}, the relation \eqref{Difference-Matrix-Elements} gives
\begin{multline}
(1-q^{n})\;K_{n}^{\text{Qtm}}\left(q^{-x};\frac{1}{\theta^2 q^{N}},N;q\right)=(1-q^{x})(1-\theta^2q^{x-1})\;K_{n}^{\text{Qtm}}\left(q^{-x};\frac{1}{\theta^2 q^{N}},N;q\right)
\\
-(1-q^{x})(1-\theta^2 q^{x-1})\;K_{n}^{\text{Qtm}}\left(q^{-(x-1)};\frac{1}{\theta^2 q^{N}},N;q\right)
\\
-\theta^2 q^{2x}(1-q^{N-x})\;K_{n}^{\text{Qtm}}\left(q^{-(x+1)};\frac{1}{\theta^2 q^{N}},N;q\right)+\theta^2 q^{2x}(1-q^{N-x})\;K_{n}^{\text{Qtm}}\left(q^{-x};\frac{1}{\theta^2 q^{N}},N;q\right),
\end{multline}
which can be seen to coincide with the one given in \cite{Koekoek-2010}.
\section{Duality relation with affine $q$-Krawtchouk polynomials}
In the preceding sections, the properties of the matrix elements $\chi_{n,x}^{(N)}$ of the unitary $q$-operator \eqref{Main-Operator} have been derived algebraically. Through the explicit expression \eqref{Matrix-Elements-Explicit} of the matrix elements in terms of the quantum $q$-Krawtchouk polynomials, the properties of these polynomials have been obtained. It is possible to express the matrix elements $\chi_{n,x}^{(N)}$ in terms of another family of orthogonal functions: the affine $q$-Krawtchouk polynomials. These polynomials are defined as \cite{Koekoek-2010}
\begin{align}
\label{Def-Aff}
K_{n}^{\text{Aff}}(q^{-x};p,N;q)=\pfq{3}{2}{q^{-n},0,q^{-x}}{pq,q^{-N}}{q,q}=\frac{(-pq)^{n}q^{\binom{n}{2}}}{(pq;q)_{n}}\;\pfq{2}{1}{q^{-n},q^{x-N}}{q^{-N}}{q,\frac{q^{-x}}{p}}.
\end{align}
By inspection of the hypergeometric formula \eqref{Formula-1} for the matrix elements, it is easily seen comparing with \eqref{Def-Aff} that they can be written as
\begin{align}
\label{Matrix-Elements-Aff}
\chi_{n,x}^{(N)}=\theta^{n-x}\qbinom{N}{x}^{1/2}\qbinom{N}{n}^{1/2}(\theta^2;q)_{x}^{1/2}(\theta^2;q)_{N-n}^{1/2}\;K_{x}^{\text{Aff}}\left(q^{-(N-n)};\frac{\theta^2}{q},N;q\right).
\end{align}
Note that here $x$ appears as the degree. The properties of the affine $q$-Krawtchouk polynomials can be obtained from those of the matrix elements. For example, with the help of the identification \eqref{Matrix-Elements-Aff}, the generating relation  \eqref{Gen-2} gives a generating function for the affine $q$-Krawtchouk polynomials
\begin{align}
\label{Gen-3}
(-p\,q\,q^{-N}v;q)_{N}\;\pfq{2}{1}{q^{-x},0}{-\frac{1}{p v}}{q,\frac{q^{x-N}}{p}}=\sum_{n=0}^{N}\frac{(q^{-N};q)_{n}(p\,q;q)_{n}}{(q;q)_{n}}q^{-\binom{n}{2}}\;K_{n}^{\text{Aff}}(q^{-(N-x)};p,N;q)\;v^{n}.
\end{align}
The RHS of \eqref{Gen-3} corresponds to one of the generating functions for the affine $q$-Krawtchouk polynomials given in \cite{Koekoek-2010}. However in the latter, the LHS is expressed in terms of a ${}_{2}\phi_{0}$ basic hypergeometric series. The two expressions can be reconciled as follows. Consider the transformation identity
\begin{align*}
\pfq{2}{1}{q^{-n},b}{c}{q,z}=\frac{b^{n}(c/b;q)_{n}}{(c;q)_{n}}\;\pfq{3}{1}{q^{-n},b,q/z}{b q^{1-n}/c}{q,\frac{z}{c}},
\end{align*}
given in Appendix III of \cite{Gasper-2004}. Taking the limit as $b\rightarrow 0$ in the above, one finds
\begin{align*}
\pfq{2}{1}{q^{-n},0}{c}{q,z}=\frac{(-c)^{n}q^{\binom{n}{2}}}{(c;q)_{n}}\;\pfq{2}{0}{q^{-n},q/z}{-}{q,\frac{z}{c}}.
\end{align*}
Using the above relation in the LHS of \eqref{Gen-3}, one easily finds
\begin{multline*}
(-p v q^{1-N};q)_{N-x}\;\pfq{2}{0}{q^{-x},p q^{N-x+1}}{-}{q,-v q^{-(N-x)}}
\\
=\sum_{n=0}^{N}\frac{(q^{-N};q)_{n}(p\,q;q)_{n}}{(q;q)_{n}}q^{-\binom{n}{2}}\;K_{n}^{\text{Aff}}(q^{-(N-x)};p,N;q)\;v^{n},
\end{multline*}
which coincides with the generating function given in \cite{Koekoek-2010}. The expression \eqref{Matrix-Elements-Aff} also implies a duality relation between the affine and the quantum $q$-Krawtchouk polynomials. Indeed, comparing \eqref{Matrix-Elements-Aff} with \eqref{Matrix-Elements-Explicit}, it follows that
\begin{align}
\label{Duality-New}
K_{n}^{\text{Qtm}}\left(q^{-x};\frac{1}{\theta^2 q^{N}},N;q\right)=\frac{(-1)^{x}(\theta^2;q)_{x}}{\theta^{2x}q^{\binom{x}{2}}}\;K_{x}^{\text{Aff}}\left(q^{-(N-n)};\frac{\theta^2}{q},N;q\right).
\end{align}
Another duality exists between the quantum and the affine $q$-Krawtchouk polynomials. This duality relation is of the form \cite{Koekoek-2010}
\begin{align*}
K_{n}^{\text{Qtm}}\left(q^{-x};p,N;q^{-1}\right)=(p^{-1}q;q)_{n}\left(-\frac{p}{q}\right)^{n}q^{-n(n-1)/2}\;K_{n}^{\text{Aff}}\left(q^{x-N};p^{-1},N;q\right).
\end{align*}
In view of the above, one could also take $q\rightarrow q^{-1}$ in every formula to have the matrix elements of the $q^{-1}$-rotation operator \eqref{Main-Operator} in terms of the affine $q$-Krawtchouk polynomials.
\section{Conclusion}
In this paper, it was shown that the quantum $q$-Krawtchouk polynomials arise as the matrix elements of unitary $q$-rotation operators expressed as $q$-exponentials in the $\mathcal{U}_{q}(sl_2)$ generators in the Schwinger realization. This algebraic interpretation was used to provide a full characterization of these orthogonal functions, as well as of the affine $q$-Krawtchouk polynomials. We now plan to use the results obtained in this paper to arrive at an algebraic characterization of the multivariate quantum (and affine) $q$-Krawtchouk polynomials introduced by Gasper and Rahman \cite{Gasper-2007}.
\section*{Acknowledgments}
L.V. wishes to acknowledge the hospitality of the Shanghai Jiao Tong University where this research project was initiated. This work much benefited also from a visit of V.X.G and L.V to the University of Hawai'i whose support we similarly wish to gratefully mention. V.X.G. holds an Alexander-Graham-Bell fellowship from the Natural Sciences and Engineering Research Council of Canada (NSERC). The research of L.V. is supported in part by NSERC.
\footnotesize
\begin{multicols}{2}

\end{multicols}
\end{document}